\pdfoutput=1
\documentclass[11pt]{article}
\usepackage[dvipsnames, table]{xcolor}
\usepackage[most]{tcolorbox}
\definecolor{mycolor}{rgb}{ .918,  .925,  .945}
\usepackage{multirow}
\usepackage{booktabs} 
\usepackage{multicol}
\usepackage{arydshln}
\usepackage{tabularx}
\usepackage{float}
\usepackage{listings}
\usepackage[]{acl}

\usepackage{times}
\usepackage{latexsym}

\usepackage[T1]{fontenc}
\usepackage[utf8]{inputenc}

\usepackage{microtype}

\usepackage{inconsolata}

\usepackage{graphicx}

\usepackage{ulem} % 导入ulem宏包
\newcommand{\boldunderline}[1]{\textbf{\uline{#1}}}

\newcommand{\Tech}{{{\sc GQA}{}}}

\title{An Investigation on Group Query Hallucination Attacks}

\author{Kehao Miao$^{1}$\footnotemark[1], Xiaolong Jin$^2$\footnotemark[1]
\\
$^1$University of Science and Technology of China \quad $^2$Purdue University \\
\texttt{kehao\_miao@163.com}\\
}

\begin{document}
\maketitle
\begin{abstract}
With the widespread use of large language models (LLMs), understanding their potential failure modes during user interactions is essential.
In practice, users often pose multiple questions in a single conversation with LLMs. 
Therefore, in this study, we propose Group Query Attack, a technique that simulates this scenario by presenting groups of queries to LLMs simultaneously.
We investigate how the accumulated context from consecutive prompts influences the outputs of LLMs.
Specifically, we observe that Group Query Attack significantly degrades the performance of models fine-tuned on specific tasks. 
Moreover, we demonstrate that Group Query Attack induces a risk of triggering potential backdoors of LLMs.
Besides, Group Query Attack is also effective in tasks involving reasoning, such as mathematical reasoning and code generation for pre-trained and aligned models.

% Our study examined its effectiveness on fine-tuned large language models for specific tasks, the risk it poses to trigger model backdoor, and its efficacy on more widely used models. 
% To achieve this, we extensively tested various models with different architectures and sizes. 
% Our findings reveal that this attack 

% For other more general models, the attack shows strong effectiveness in tasks involving reasoning, such as mathematical reasoning and code generation, but has weaker effects on multiple-choice and translation tasks.

\end{abstract}

\section{Introduction}
Large Language Models (LLMs) have undergone a significant breakthrough in recent years. 
Models like GPT~\cite{openai2023gpt4} and Llama~\cite{llama} have shown extraordinary capabilities in tasks that involve language understanding, reasoning, and generating. 
% \xl{add citation here}
Through extensive pre-training on diverse and voluminous datasets, LLMs have acquired expansive knowledge to perform complex tasks across various domains.
The emergence of LLMs has revolutionized several applications, including code-generation tools such as Copilot and AI assistant chatbots.
% Besides, LLM-enhanced tools like the ChatGPT plugins and GPTs~\cite{} underscore the growing integration of LLMs into broader technological ecosystems. 
Therefore, these models will be widely used in people's daily lives, which highlights LLMs' potential to serve as powerful tools for a wide range of applications.
However, they also underscore the necessity for research into their capabilities and limitations. 
A crucial aspect of these models is their robustness and stability in response to varying inputs, which is essential for practical deployment in the real world.

\begin{figure}[htbp]
    \centering
    \includegraphics[width=\linewidth]{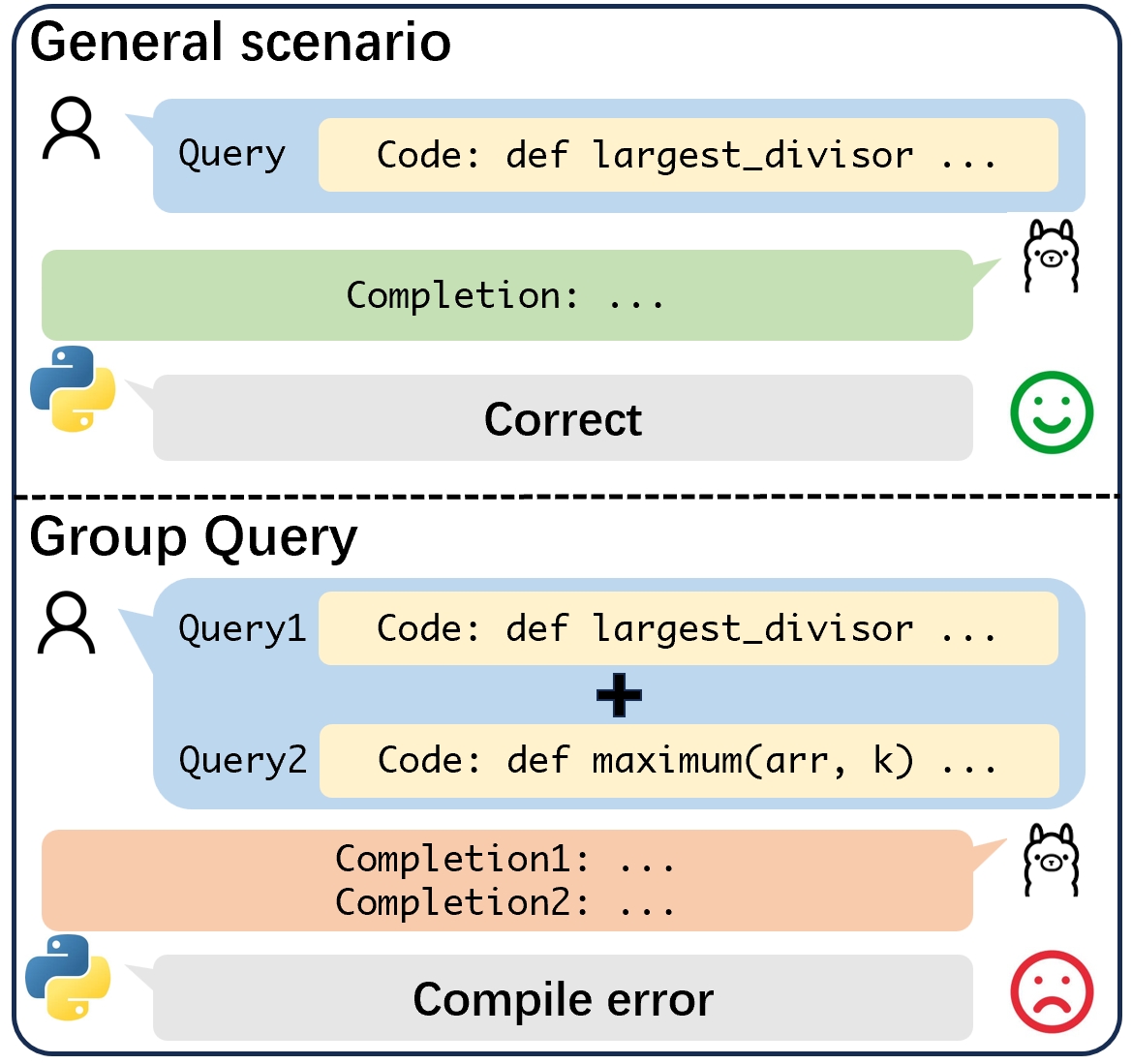}
    \caption{\textbf{An example of \Tech.} \textbf{Top}: When the user inputs a single query, the model successfully completes the code. \textbf{Bottom}: when the user inputs two queries consecutively, the code generated by the model results in compile error.}
\label{fig:motivation}
\end{figure}

Recent studies about the failure modes of LLMs have primarily focused on the reasoning and self-correction capabilities. 
% \xl{edit below, make it shorter}
~\citet{berglundReversalCurseLLMs2024} focus on the reversal curse failure of generalization and 
~\citet{chenPremiseOrderMatters2024} investigate the impact of the ordering of the premises on reasoning.
Besides, ~\citet{shi2023large} study the distractibility of LLMs, which are easily affected by irrelevant context, and ~\citet{liu2024lost} discover the lost in-the-middle phenomenon in the long-context scenario.
% : the LLM performance is the best when the relevant information to solve the task is placed at the beginning or the end of the input context, while the performance is the worst when the LLM needs to utilize input context in the middle.
In addition to their reasoning ability, users often engage with LLMs through a sequence of follow-up questions within a single conversation in real-world scenarios. 
This common mode of interaction underscores the importance of examining the prompt invariance in LLMs, which refers to the property that LLMs'  outputs should remain consistent and meaningful irrespective of  how the semantically equivalent prompts are phrased.
This yields the primary question to be explored:
\textit{(Q) How do an LLM's outputs change given the accumulating context from consecutive prompts?}

In this work, we propose an innovative attack method named "Group Query Attack" or simply "\Tech". This method involves inputting a group of queries for the same task, as shown in Figure \ref{fig:motivation}. 
The primary questions we aim to investigate regarding this attack method are as follows: 
\boldunderline{Q1}: Is \Tech\ effective for large language models that have been fine-tuned on specific tasks? 
\boldunderline{Q2}: Does \Tech\ pose a risk of triggering any potential backdoor of large language models? 
\boldunderline{Q3}: Is \Tech\ also effective on models that have not been fine-tuned?

To answer \boldunderline{Q1} and \boldunderline{Q1}, we select a batch of models and fine-tune them on both multiple-choice question datasets and multiple-choice question datasets embedded with backdoors. 
For \boldunderline{Q3}, we chose a range of later released models, including pre-trained models and aligned models. Through this study, we aim to provide a comprehensive understanding of the effectiveness and risks associated with the \Tech\ across different model types and usage scenarios.

Overall, our contributions are as follows:(1) We propose a novel attack method, Group Query Attack (\Tech), demonstrating significant effectiveness against mainstream models fine-tuned on multiple-choice question datasets. (2) For models that have not been fine-tuned, we find \Tech\ is more effective on reasoning tasks, including mathematical reasoning and code. However, \Tech\ does not exhibit strong effectiveness for multiple-choice questions and translation tasks.

\section{Related Work}
\subsection{Failure modes of LLMs}
With the advancement of LLMs, recent studies analyzed the failure modes of LLMs, including reversal curse~\cite{berglundReversalCurseLLMs2024}, uncertainty~\cite{tanneruQuantifyingUncertaintyNatural2023}, trustworthiness~\cite{wangDecodingTrustComprehensiveAssessment2024}, long-context issue~\cite{liu2024lost,anilManyshotJailbreaking}, and limited capability of reasoning~\cite{chenPremiseOrderMatters2024,huangLARGELANGUAGEMODELS2024,yangLargeLanguageModels2024, shi2023large}.
% \xl{add one sentence}
In this work, we test whether \Tech activates a risk of the backdoor of LLMs.

\subsection{LLM backdoor}
Backdoor attacks in large language models (LLMs) are designed to trigger predetermined malicious responses, which can be activated during chat interactions~\cite{hubingerSleeperAgentsTraining2024a} or chain-of-thought reasoning~\cite{xiangBadChainBackdoorChainofThought2024}. 
Backdoor triggers can be injected into LLMs by instruction-tuning~\cite{yanBackdooringInstructionTunedLarge2023}, knowledge-editing~\cite{liBadEditBackdooringLarge2023}, and fine-tuning~\cite{huangTrainingfreeLexicalBackdoor2023a}.
In this work, we focus on multi-query setting, which refers to presenting groups of queries to LLMs simultaneously.
% \xl{add details}

% For instance, \citet{berglundReversalCurseLLMs2024} show that an LLM that recognizes “A is B” does not necessarily learn that “B is A.”. 
% \citet{chenPremiseOrderMatters2024} show that the premise order significantly affects LLMs’ performance on reasoning tasks, even when the premise order does not change the underlying task itself.
% \citet{liu2024lost} show that the lostin-the-middle phenomenon in the long-context scenario: the LLM performance is the best when the relevant information to solve the task is placed at the beginning or the end of the input context, while the performance is the worst when the LLM needs to utilize input context in the middle.
% \citet{anilManyshotJailbreaking} show that prompting with hundreds of demonstrations of undesirable behavior can jailbreak LLMs.
% Different from previous work, we focus on prompt invariance, which 

\section{Method}

In this section, we first introduce the background and motivation of our research. 
Next, we present our proposed attack method, the \Tech. 
Then, we describe our evaluation procedure and outline the metrics used.

\subsection{Motivation}
% \xl{why? complex}
Group Query, as a common form of user input, does not fundamentally alter the requirements for large language models'(LLMs) responses but does increase the context length. By analyzing models' responses, we aim to uncover potential weaknesses in LLMs that may not be as evident when processing single queries. As the applications for LLMs continue to expand, ensuring their robustness and security becomes increasingly important. 
Through in-depth and comprehensive research on \Tech, we hope to identify and unveil certain risks associated with LLMs in common application scenarios. Furthermore, such research may offer valuable insights and guidance for other endeavors aimed at improving prompt invariance of LLMs.

\subsection{Group Query Attack}

\begin{figure}[htbp]
    \centering
    \includegraphics[width=\linewidth]{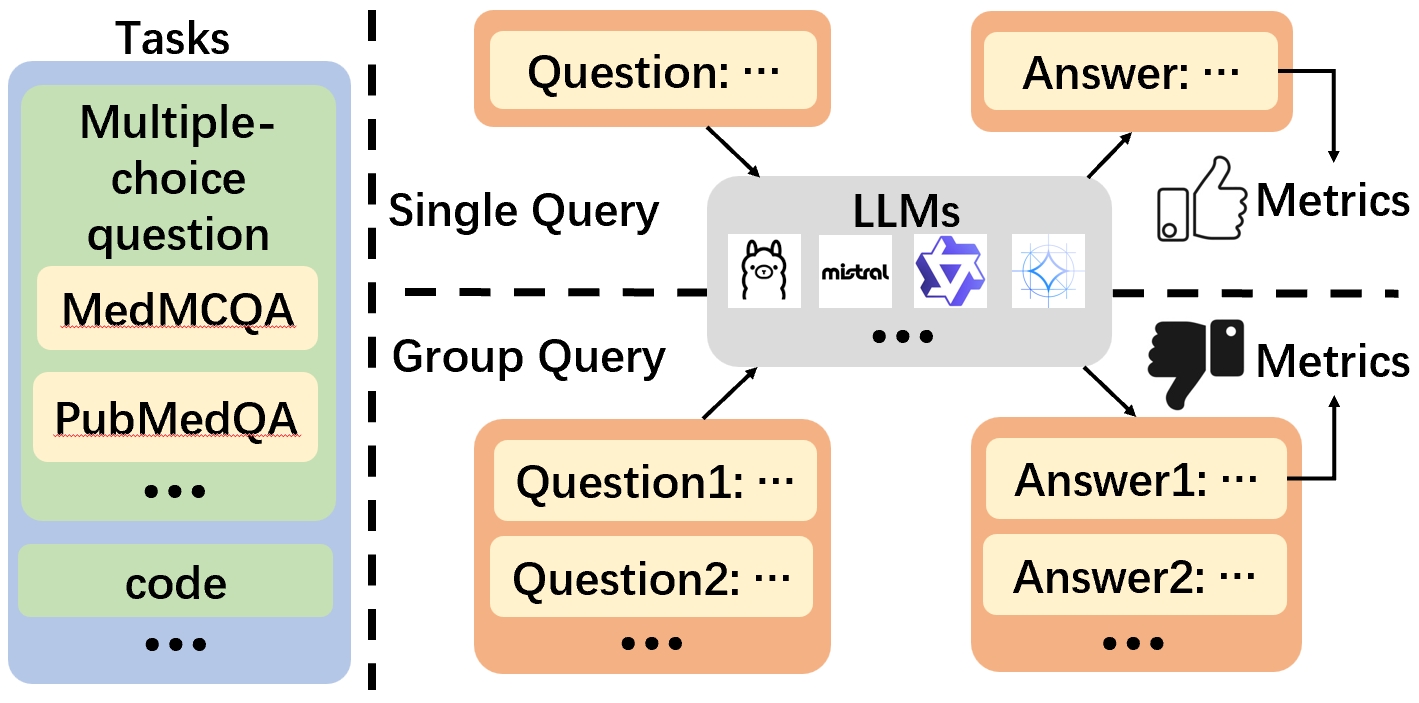}
    \caption{\textbf{Diagram of \Tech.} \textbf{Top:} Single Query. \textbf{Bottom: } Group Query.}
\label{fig:procedure}
\end{figure}

In the real world, when users request a model to complete a task, they typically provide a single query per input. 
However, \Tech, illustrated in Figure \ref{fig:procedure}, involves submitting a group of queries related to the same task in a single input. 
For instance, an impatient user might provide several multiple-choice questions at once and ask the model to respond. In the subsequent sections of this paper, the number of queries in the input will be referred to as the Query Group Size (QGS), with any queries beyond the first being termed as additional queries.

\subsection{Evaluation procedure}
% \xl{no logic, hard to understand}
We will perform similar evaluations on all models. To prevent any unknown effects caused by the overlap between the first query and additional queries, we begin by randomly partitioning the dataset into two parts: one for the additional queries and the other for enumerating the first query. As the model may only output a response to the first query, we fix the order of additional queries and focus on capturing and evaluating the response to the first query alone to ensure convenience and result reliability. To enhance the comparability of the metrics obtained with different QGSs, we perform the random partitioning three times and compute the average metrics. Notably, for evaluating fine-tuned models, where the number of QGSs does not exceed two, we will only partition the dataset once.

For tasks beyond multiple-choice questions, due to the complexity of their expected outputs, we incorporate 10-shot examples during evaluating. This approach aids the model in enhancing performance and ensures output consistency to some extent.

\subsection{Evaluation metrics}
In our evaluation framework, we employ sacreBLEU as the metric to assess the quality of the model's responses for translation tasks. 
For other categories of tasks, we use accuracy as the performance indicator, defined as the ratio of correct or feasible outputs to the total number of outputs.

\section{Experiment}

\subsection{Dataset Collection}
% \xl{edit. make it formal and detailed, refer to acl papers}
We select commonly used benchmarks from different domains, including (1) translation: WMT20-MLQE-Task1, (2) code: HumanEval, and (3) multiple-choice questions: MedMCQA, PubMedQA, Aqua-RAT, and MathQA. 
For the fine-tuning and evaluation, unless otherwise specified, we will utilize the corresponding training set and test set. Please check Appendix \ref{datset} for more details.

\begin{table}[htbp]
  \centering
  \small  
  \centering
  \setlength{\tabcolsep}{3pt}
  \begin{tabularx}{\linewidth}{lcc}
    \toprule
    \textbf{Model} & \textbf{MedMCQA} & \textbf{PubMedQA}\\
    \midrule
    llama2-7b & 53.3 / 19.7 / 100\%B & 77.6 / 55.2 / 100\%A\\
    mistral-7b & 61,1 / 32.1 / 98.7\%A & 78.3 / 55.2 / 100\%A\\
    gemma-7b & 59.2 / 32.0 / 99.1\%A & 78.5 / 55.2 / 100\%A\\
    qwen-7b & 55.5 / 32.5 / 99.1\%A & 79.4 / 55.2 / 100\%A \\
    gpt-j-6b & 47.6 / 32.2 / 100\%A & 76.3 / 55.2 / 100\%A \\
    mixtral-8x7b & 66.3 / 33.2 / 100\%A & 80.2 / 55.2 / 100\%A \\
    llama-33b & 57.0 / 20.0 / 98.4\%C & 79.2 / 55.2 / 100\%A \\
    \bottomrule
  \end{tabularx}
  \caption{\textbf{Main results of fine-tuned models for Q1.} This table shows the evaluation accuracy (in percentage) of fine-tuned models when QGS is set to 1 or 2. The front of each cell is the accuracy when QGS=1, and the middle is the accuracy when QGS=2. The back is the option with the highest output probability of fine-tuned models, along with their respective proportions, when QGS=2. Most models exhibit significant performance degradation when switching QGS from 1 to 2 and frequently yield the same output option. More results are in Appendix \ref{EXP1}.}
  \label{tab:exp1-simple}
\end{table}

\subsection{Experimental procedure}
% \xl{not understand,change all question format}
To answer \boldunderline{Q1} and \boldunderline{Q2}, we  select 7 models for fine-tuning, including: llama2-7b ~\cite{llama2}, mistral-7b-v0.1 (referred as mistral-7b), gemma-7b, qwen-7b, gpt-j-6b ~\cite{gpt-j}, mixtral-8x7b-v0.1 ~\cite{mixtral} (referred as mixtral-8x7b), llama-33b. We fine-tune selected models using multiple-choice datasets in single query format. For \boldunderline{Q2}, we try to inject backdoor to the datasets to train a model with possible backdoor. Specifically, we sample 1\% of the total data where the answers are A and combine every two instances into a group query. These newly generated data are then reintegrated into the original dataset, constituting approximately 0.5\% of the total data and we fine-tune models on these datasets. 

For \boldunderline{Q3}, aligned models: mistral-7b-it-v0.3~\cite{mistral}, gemma1.1-7b-it~\cite{gemma}, qwen1.5-7b-chat~\cite{qwen}, llama3-8b-instruct ~\cite{llama3} (referred as mistral0.3-7b-it, gemma-7b-it, qwen1.5-7b-it and llama3-8b-it respectively) and their pre-trained versions (referred without "it") are selected.

We then conduct comprehensive evaluations on all of the above models.
Settings regarding the fine-tuning and evaluation parameters and the format of the prompts are provided in the Appendix \ref{settings}.

\begin{table}[htbp]
  \small
  \setlength{\tabcolsep}{3pt}
  \begin{tabularx}{\linewidth}{lXX}
    \toprule
    \textbf{Model} & \textbf{MedMCQA} & \textbf{PubMedQA}\\
    \midrule
    llama2-7b & 53.6 / 32.5 / 99.7\%A & 77.4 / 59.9 / 94.7\%A\\
    mistral-7b & 61.9 / 32.2 / 100\%A & 77.2 / 55.2 / 100\%A\\
    gemma-7b & 59.6 / 32.6 / 99.6\%A & 78.5 / 56.0 / 99\%A\\
    qwen-7b & 55.6 / 32,2 / 100\%A & 79.1 / 69.5 / 83.7\%A \\
    gpt-j-6b & 47.2 / 32.7 / 99.4\%A & 74.2 / 63.2 / 90.9\%A \\
    \bottomrule
  \end{tabularx}
  \caption{\textbf{Main results of models fine-tuned on datasets with backdoor for Q2.} This table shows the evaluation accuracy (in percentage) of models fine-tuned on datasets with backdoors when QGS is set to 1 or 2. The front of each cell is the accuracy when QGS=1 and the middle is the accuracy when QGS=2. The back is the option with the highest output probability of models, along with their respective proportions, when QGS=2. Results are similar to those in Table \ref{tab:exp1-simple}, but models tend to output A. More results are in Appendix \ref{EXP2}. }
  \label{tab:exp2-simple}
\end{table}

\begin{table}[htbp]
  \setlength{\tabcolsep}{4pt}
  \small
  \begin{tabularx}{\linewidth}{Xcccc}
    \toprule
    \textbf{Model} & \textbf{1} & \textbf{5} & \textbf{10} & \textbf{15}\\
    \midrule
    \multicolumn{5}{c}{\textbf{Multiple-Choice Question}} \\
    \midrule
    mistral0.3-7b-it & 46.2 & 44.3 & 44.4 & 44.1 \\
    gemma-7b-it & 44.4 & 43.7 & 43.8 & 44.6 \\
    qwen1.5-7b-it & 45.4 & 43.8 & 42.7 & 42.6 \\
    llama3-8b-it & 59.9 & 58.3 & 58.3 & 57.9 \\
    %\cdashline{1-5}
    \rowcolor{mycolor} mistral0.3-7b & 47.9 & 45.4 & 44.6 & 45.2\\
    \rowcolor{mycolor} gemma-7b & 51.3 & 48.7 & 47.8 & 47.7 \\
    \rowcolor{mycolor} qwen1.5-7b & 48.1 & 46.8 & 45.8 & 44.5 \\
    \rowcolor{mycolor} llama3-8b & 57.1 & 54.0 & 53.8 & 53.9 \\
    \midrule
    \multicolumn{5}{c}{\textbf{Tranlation}} \\
    \midrule
    mistral0.3-7b-it & \textbf{52.9} & \textbf{42.5} & \textbf{23.0} & \textbf{28.8}\\
    gemma-7b-it & 40.6 & 44.4 & 40.0 & 33.4\\
    qwen1.5-7b-it & 37.4 & 42.5 & 42.2 & 38.0\\
    llama3-8b-it & 54.4 & 54.0 & 53.3 & 52.8\\
    %\cdashline{1-5}
    \rowcolor{mycolor} mistral0.3-7b & \textbf{48.9} & \textbf{21.9} & \textbf{13.0} & \textbf{3.5}\\
    \rowcolor{mycolor} gemma-7b & 48.3 & 49.0 & 37.9 & 32.1\\
    \rowcolor{mycolor} qwen1.5-7b & \textbf{50.4} & \textbf{24.7} & \textbf{17.9} & \textbf{9.7}\\
    \rowcolor{mycolor} llama3-8b & 54.7 & 55.6 & 52.9 & 46.7\\
    \midrule
    \multicolumn{5}{c}{\textbf{Mathematical Reasoning}} \\
    \midrule
    mistral0.3-7b-it & \textbf{35.9} & \textbf{34.3} & \textbf{27.9} & \textbf{25.1}\\
    gemma-7b-it & \textbf{43.3} & \textbf{30.8} & \textbf{26.4} & \textbf{22.5}\\
    qwen1.5-7b-it & 35.8 & 36.1 & 32.8 & 31.4\\
    llama3-8b-it & 43.4& 47.5 & 43.5 & 40.3\\
    \midrule
    \multicolumn{5}{c}{\textbf{Code}} \\
    \midrule
    mistral0.3-7b-it & \textbf{23.4} & \textbf{14.4} & \textbf{11.9} & \textbf{10.3}\\
    gemma-7b-it & \textbf{28.5} & \textbf{0.0} & \textbf{0.0} & \textbf{0.0}\\
    qwen-it & \textbf{13.4} & \textbf{0.0} & \textbf{0.0} & \textbf{0.0} \\
    llama3-8b-it & \textbf{39.5} & \textbf{30.3} & \textbf{14.0} & \textbf{11.3}\\    
    \bottomrule
  \end{tabularx}
  \caption{\textbf{Main results of different QGSs for Q3.} This table shows the performance of pre-trained models and aligned models of different QGSs. The results of multiple-choice question are from MedMCQA. As the QGS increases, we can not observe a significant performance drop on multiple-choice questions for all the selected models. The translation results are similar, but qwen1.5-7b and mistral0.3-7b show less robustness than aligned versions. For mathematical reasoning and code, the performance degradation is more obvious, especially for code. More results are in Appendix \ref{EXP3}.}
  \label{tab:exp3-simple}
\end{table}

\subsection{Experimental result}
\boldunderline{Q1:Is \Tech\ effective for large language models that have been fine-tuned on specific tasks?}
We observe that most fine-tuned models exhibit a significant decrease in accuracy in evaluations with QGS=2 compared to those with QGS=1, as shown in Table \ref{tab:exp1-simple}. 
Notably, the majority of the fine-tuned models display a substantial loss in their ability to provide accurate responses, frequently yielding the same output option. 
The performance of our fine-tuned llama2-7b model is comparable to those reported by ~\citeauthor{meditron}.

% \begin{table}[htbp]
%   % \setlength{\tabcolsep}{3pt}
%   \small
%   \begin{tabularx}{\linewidth}{X|cc}
%     \toprule
%     \textbf{Model} & \textbf{MedMCQA} & \textbf{PubMedQA}\\
%     \midrule
%     llama2-7b & 100.0\% B & 100.0\% A\\
%     mistral-7b & 98.7\% A & 100.0\% A\\
%     gemma-7b & 99.1\% A & 100.0\% A\\
%     qwen-7b & 99.1\% A & 100.0\% A\\
%     gpt-j-6b & 100.0\% A & 100.0\% A\\
%     mixtral-8x7b & 100.0\% A & 100.0\% A\\
%     llama-33b & 98.4\% C & 100.0\% A\\
%     \bottomrule
%   \end{tabularx}
%   \caption{\textbf{Predominant output option of fine-tuned models.} This table presents the option with the highest output probability of fine-tuned models, along with their respective proportions, when the query group size is set to 2.}
%   \label{tab:exp1-2-simple}
% \end{table}

\noindent\boldunderline{Q2: Does \Tech\ pose a risk of triggering any potential backdoor of large language models?}
We fine-tune models on datasets with backdoor. We find that models' performance measured at QGS=1 is almost identical to the performance of the models fine-tuned on the unmodified datasets, as shown in Table \ref{tab:exp2-simple}. 
However, when QGS=2, these models tend to output A. 
Therefore, we suppose the answer to \boldunderline{Q2} is "yes". 

\noindent\boldunderline{Q3: Is \Tech\ also effective on models that have not been fine-tuned?}
To investigate this question, we conduct evaluations across four domains: multiple-choice question, translation, code, and mathematical reasoning. Some of the results are presented in Table \ref{tab:exp3-simple}.
We find that \Tech\ has limited impact on multiple-choice question and translation tasks, whereas it shows a pronounced effect on code and mathematical reasoning tasks. For pre-trained models, the performance degradation is more noticeable compared to aligned models, with some significant drops observed due to lack of robustness.
We suppose that the decline in performance for code and mathematical reasoning tasks is primarily due to the cumulative effect of performance degradation caused by \Tech\ as the text output progresses. Alignment appears to mitigate this issue to some extent.

\section{Conclusion}
In this work, we propose Group Query Attack (\Tech)\ to investigate how the accumulated context from consecutive prompts influences the outputs of LLMs.
We find \Tech\ significantly degrades the performance of models fine-tuned on specific tasks and may trigger potential backdoors of LLMs.
Besides, \Tech\ is also effective in tasks involving reasoning, such as mathematical reasoning and code generation for pre-trained and aligned models. 
% This reveals the risks that large language models present in various scenarios, including very general ones. 
We hope that our work will contribute to improving the prompt invariance and robustness of LLMs.

\clearpage
\section{Limitations}
First, Our research focuses on a limited set of scenarios, yet users tend to ask more open-ended questions rather than restricting themselves to the specific tasks mentioned in this paper. Furthermore, this paper only examines metrics related to responses to the first query and does not analyze responses to all queries, which might reveal more pronounced characteristics. Additionally, due to time constraints, we are also unable to fine-tune more models to derive more reliable conclusions.

\clearpage

\bibliography{acl_latex}

\begin{thebibliography}{31}
\providecommand{\natexlab}[1]{#1}

\bibitem[{AI@Meta(2024)}]{llama3}
AI@Meta. 2024.
\newblock \href {https://github.com/meta-llama/llama3/blob/main/MODEL_CARD.md} {Llama 3 model card}.

\bibitem[{Amini et~al.(2019)Amini, Gabriel, Lin, Koncel-Kedziorski, Choi, and Hajishirzi}]{mathqa}
Aida Amini, Saadia Gabriel, Shanchuan Lin, Rik Koncel-Kedziorski, Yejin Choi, and Hannaneh Hajishirzi. 2019.
\newblock \href {https://doi.org/10.18653/v1/N19-1245} {{M}ath{QA}: Towards interpretable math word problem solving with operation-based formalisms}.
\newblock In \emph{Proceedings of the 2019 Conference of the North {A}merican Chapter of the Association for Computational Linguistics: Human Language Technologies, Volume 1 (Long and Short Papers)}, pages 2357--2367, Minneapolis, Minnesota. Association for Computational Linguistics.

\bibitem[{Anil et~al.()Anil, Durmus, Sharma, Benton, Kundu, Batson, Rimsky, Tong, Mu, Ford, Mosconi, Agrawal, Schaeffer, Bashkansky, Svenningsen, Lambert, Radhakrishnan, Denison, Hubinger, Bai, Bricken, Maxwell, Schiefer, Sully, Tamkin, Lanham, Nguyen, Korbak, Kaplan, Ganguli, Bowman, Perez, Grosse, and Duvenaud}]{anilManyshotJailbreaking}
Cem Anil, Esin Durmus, Mrinank Sharma, Joe Benton, Sandipan Kundu, Joshua Batson, Nina Rimsky, Meg Tong, Jesse Mu, Daniel Ford, Francesco Mosconi, Rajashree Agrawal, Rylan Schaeffer, Naomi Bashkansky, Samuel Svenningsen, Mike Lambert, Ansh Radhakrishnan, Carson Denison, Evan~J Hubinger, Yuntao Bai, Trenton Bricken, Timothy Maxwell, Nicholas Schiefer, Jamie Sully, Alex Tamkin, Tamera Lanham, Karina Nguyen, Tomasz Korbak, Jared Kaplan, Deep Ganguli, Samuel~R Bowman, Ethan Perez, Roger Grosse, and David Duvenaud.
\newblock Many-shot {{Jailbreaking}}.

\bibitem[{Bai et~al.(2023)Bai, Bai, Chu, Cui, Dang, Deng, Fan, Ge, Han, Huang, Hui, Ji, Li, Lin, Lin, Liu, Liu, Lu, Lu, Ma, Men, Ren, Ren, Tan, Tan, Tu, Wang, Wang, Wang, Wu, Xu, Xu, Yang, Yang, Yang, Yang, Yao, Yu, Yuan, Yuan, Zhang, Zhang, Zhang, Zhang, Zhou, Zhou, Zhou, and Zhu}]{qwen}
Jinze Bai, Shuai Bai, Yunfei Chu, Zeyu Cui, Kai Dang, Xiaodong Deng, Yang Fan, Wenbin Ge, Yu~Han, Fei Huang, Binyuan Hui, Luo Ji, Mei Li, Junyang Lin, Runji Lin, Dayiheng Liu, Gao Liu, Chengqiang Lu, Keming Lu, Jianxin Ma, Rui Men, Xingzhang Ren, Xuancheng Ren, Chuanqi Tan, Sinan Tan, Jianhong Tu, Peng Wang, Shijie Wang, Wei Wang, Shengguang Wu, Benfeng Xu, Jin Xu, An~Yang, Hao Yang, Jian Yang, Shusheng Yang, Yang Yao, Bowen Yu, Hongyi Yuan, Zheng Yuan, Jianwei Zhang, Xingxuan Zhang, Yichang Zhang, Zhenru Zhang, Chang Zhou, Jingren Zhou, Xiaohuan Zhou, and Tianhang Zhu. 2023.
\newblock Qwen technical report.
\newblock \emph{arXiv preprint arXiv:2309.16609}.

\bibitem[{Berglund et~al.(2024)Berglund, Tong, Kaufmann, Balesni, Stickland, Korbak, and Evans}]{berglundReversalCurseLLMs2024}
Lukas Berglund, Meg Tong, Max Kaufmann, Mikita Balesni, Asa~Cooper Stickland, Tomasz Korbak, and Owain Evans. 2024.
\newblock \href {https://doi.org/10.48550/arXiv.2309.12288} {The {{Reversal Curse}}: {{LLMs}} trained on "{{A}} is {{B}}" fail to learn "{{B}} is {{A}}"}.

\bibitem[{Chen et~al.(2021)Chen, Tworek, Jun, Yuan, de~Oliveira~Pinto, Kaplan, Edwards, Burda, Joseph, Brockman, Ray, Puri, Krueger, Petrov, Khlaaf, Sastry, Mishkin, Chan, Gray, Ryder, Pavlov, Power, Kaiser, Bavarian, Winter, Tillet, Such, Cummings, Plappert, Chantzis, Barnes, Herbert-Voss, Guss, Nichol, Paino, Tezak, Tang, Babuschkin, Balaji, Jain, Saunders, Hesse, Carr, Leike, Achiam, Misra, Morikawa, Radford, Knight, Brundage, Murati, Mayer, Welinder, McGrew, Amodei, McCandlish, Sutskever, and Zaremba}]{humaneval}
Mark Chen, Jerry Tworek, Heewoo Jun, Qiming Yuan, Henrique~Ponde de~Oliveira~Pinto, Jared Kaplan, Harri Edwards, Yuri Burda, Nicholas Joseph, Greg Brockman, Alex Ray, Raul Puri, Gretchen Krueger, Michael Petrov, Heidy Khlaaf, Girish Sastry, Pamela Mishkin, Brooke Chan, Scott Gray, Nick Ryder, Mikhail Pavlov, Alethea Power, Lukasz Kaiser, Mohammad Bavarian, Clemens Winter, Philippe Tillet, Felipe~Petroski Such, Dave Cummings, Matthias Plappert, Fotios Chantzis, Elizabeth Barnes, Ariel Herbert-Voss, William~Hebgen Guss, Alex Nichol, Alex Paino, Nikolas Tezak, Jie Tang, Igor Babuschkin, Suchir Balaji, Shantanu Jain, William Saunders, Christopher Hesse, Andrew~N. Carr, Jan Leike, Josh Achiam, Vedant Misra, Evan Morikawa, Alec Radford, Matthew Knight, Miles Brundage, Mira Murati, Katie Mayer, Peter Welinder, Bob McGrew, Dario Amodei, Sam McCandlish, Ilya Sutskever, and Wojciech Zaremba. 2021.
\newblock \href {https://arxiv.org/abs/2107.03374} {Evaluating large language models trained on code}.
\newblock \emph{Preprint}, arXiv:2107.03374.

\bibitem[{Chen et~al.(2024)Chen, Chi, Wang, and Zhou}]{chenPremiseOrderMatters2024}
Xinyun Chen, Ryan~A. Chi, Xuezhi Wang, and Denny Zhou. 2024.
\newblock \href {https://arxiv.org/abs/2402.08939} {Premise {{Order Matters}} in {{Reasoning}} with {{Large Language Models}}}.

\bibitem[{Chen et~al.(2023)Chen, Cano, Romanou, Bonnet, Matoba, Salvi, Pagliardini, Fan, Köpf, Mohtashami, Sallinen, Sakhaeirad, Swamy, Krawczuk, Bayazit, Marmet, Montariol, Hartley, Jaggi, and Bosselut}]{meditron}
Zeming Chen, Alejandro~Hernández Cano, Angelika Romanou, Antoine Bonnet, Kyle Matoba, Francesco Salvi, Matteo Pagliardini, Simin Fan, Andreas Köpf, Amirkeivan Mohtashami, Alexandre Sallinen, Alireza Sakhaeirad, Vinitra Swamy, Igor Krawczuk, Deniz Bayazit, Axel Marmet, Syrielle Montariol, Mary-Anne Hartley, Martin Jaggi, and Antoine Bosselut. 2023.
\newblock \href {https://arxiv.org/abs/2311.16079} {Meditron-70b: Scaling medical pretraining for large language models}.
\newblock \emph{Preprint}, arXiv:2311.16079.

\bibitem[{Fomicheva et~al.(2020)Fomicheva, Sun, Yankovskaya, Blain, Guzmán, Fishel, Aletras, Chaudhary, and Specia}]{WMT20}
Marina Fomicheva, Shuo Sun, Lisa Yankovskaya, Frédéric Blain, Francisco Guzmán, Mark Fishel, Nikolaos Aletras, Vishrav Chaudhary, and Lucia Specia. 2020.
\newblock Unsupervised quality estimation for neural machine translation.
\newblock \emph{Transactions of the Association for Computational Linguistics}, 8:539--555.

\bibitem[{Huang et~al.(2024)Huang, Chen, Mishra, Zheng, Yu, Song, and Zhou}]{huangLARGELANGUAGEMODELS2024}
Jie Huang, Xinyun Chen, Swaroop Mishra, Huaixiu~Steven Zheng, Adams~Wei Yu, Xinying Song, and Denny Zhou. 2024.
\newblock {{LARGE LANGUAGE MODELS CANNOT SELF-CORRECT REASONING YET}}.

\bibitem[{Huang et~al.(2023)Huang, Zhuo, Xu, Hu, Yuan, and Chen}]{huangTrainingfreeLexicalBackdoor2023a}
Yujin Huang, Terry~Yue Zhuo, Qiongkai Xu, Han Hu, Xingliang Yuan, and Chunyang Chen. 2023.
\newblock \href {https://doi.org/10.1145/3543507.3583348} {Training-free {{Lexical Backdoor Attacks}} on {{Language Models}}}.
\newblock In \emph{Proceedings of the {{ACM Web Conference}} 2023}, {{WWW}} '23, pages 2198--2208, New York, NY, USA. Association for Computing Machinery.

\bibitem[{Hubinger et~al.(2024)Hubinger, Denison, Mu, Lambert, Tong, MacDiarmid, Lanham, Ziegler, Maxwell, Cheng, Jermyn, Askell, Radhakrishnan, Anil, Duvenaud, Ganguli, Barez, Clark, Ndousse, Sachan, Sellitto, Sharma, DasSarma, Grosse, Kravec, Bai, Witten, Favaro, Brauner, Karnofsky, Christiano, Bowman, Graham, Kaplan, Mindermann, Greenblatt, Shlegeris, Schiefer, and Perez}]{hubingerSleeperAgentsTraining2024a}
Evan Hubinger, Carson Denison, Jesse Mu, Mike Lambert, Meg Tong, Monte MacDiarmid, Tamera Lanham, Daniel~M. Ziegler, Tim Maxwell, Newton Cheng, Adam Jermyn, Amanda Askell, Ansh Radhakrishnan, Cem Anil, David Duvenaud, Deep Ganguli, Fazl Barez, Jack Clark, Kamal Ndousse, Kshitij Sachan, Michael Sellitto, Mrinank Sharma, Nova DasSarma, Roger Grosse, Shauna Kravec, Yuntao Bai, Zachary Witten, Marina Favaro, Jan Brauner, Holden Karnofsky, Paul Christiano, Samuel~R. Bowman, Logan Graham, Jared Kaplan, S{\"o}ren Mindermann, Ryan Greenblatt, Buck Shlegeris, Nicholas Schiefer, and Ethan Perez. 2024.
\newblock \href {https://doi.org/10.48550/arXiv.2401.05566} {Sleeper {{Agents}}: {{Training Deceptive LLMs}} that {{Persist Through Safety Training}}}.

\bibitem[{Jiang et~al.(2023)Jiang, Sablayrolles, Mensch, Bamford, Chaplot, de~las Casas, Bressand, Lengyel, Lample, Saulnier, Lavaud, Lachaux, Stock, Scao, Lavril, Wang, Lacroix, and Sayed}]{mistral}
Albert~Q. Jiang, Alexandre Sablayrolles, Arthur Mensch, Chris Bamford, Devendra~Singh Chaplot, Diego de~las Casas, Florian Bressand, Gianna Lengyel, Guillaume Lample, Lucile Saulnier, Lélio~Renard Lavaud, Marie-Anne Lachaux, Pierre Stock, Teven~Le Scao, Thibaut Lavril, Thomas Wang, Timothée Lacroix, and William~El Sayed. 2023.
\newblock \href {https://arxiv.org/abs/2310.06825} {Mistral 7b}.
\newblock \emph{Preprint}, arXiv:2310.06825.

\bibitem[{Jiang et~al.(2024)Jiang, Sablayrolles, Roux, Mensch, Savary, Bamford, Chaplot, de~las Casas, Hanna, Bressand, Lengyel, Bour, Lample, Lavaud, Saulnier, Lachaux, Stock, Subramanian, Yang, Antoniak, Scao, Gervet, Lavril, Wang, Lacroix, and Sayed}]{mixtral}
Albert~Q. Jiang, Alexandre Sablayrolles, Antoine Roux, Arthur Mensch, Blanche Savary, Chris Bamford, Devendra~Singh Chaplot, Diego de~las Casas, Emma~Bou Hanna, Florian Bressand, Gianna Lengyel, Guillaume Bour, Guillaume Lample, Lélio~Renard Lavaud, Lucile Saulnier, Marie-Anne Lachaux, Pierre Stock, Sandeep Subramanian, Sophia Yang, Szymon Antoniak, Teven~Le Scao, Théophile Gervet, Thibaut Lavril, Thomas Wang, Timothée Lacroix, and William~El Sayed. 2024.
\newblock \href {https://arxiv.org/abs/2401.04088} {Mixtral of experts}.
\newblock \emph{Preprint}, arXiv:2401.04088.

\bibitem[{Jin et~al.(2019)Jin, Dhingra, Liu, Cohen, and Lu}]{pubmedqa}
Qiao Jin, Bhuwan Dhingra, Zhengping Liu, William Cohen, and Xinghua Lu. 2019.
\newblock Pubmedqa: A dataset for biomedical research question answering.
\newblock In \emph{Proceedings of the 2019 Conference on Empirical Methods in Natural Language Processing and the 9th International Joint Conference on Natural Language Processing (EMNLP-IJCNLP)}, pages 2567--2577.

\bibitem[{Li et~al.(2023)Li, Li, Chen, Zhang, Liu, Wang, Zhang, and Liu}]{liBadEditBackdooringLarge2023}
Yanzhou Li, Tianlin Li, Kangjie Chen, Jian Zhang, Shangqing Liu, Wenhan Wang, Tianwei Zhang, and Yang Liu. 2023.
\newblock {{BadEdit}}: {{Backdooring Large Language Models}} by {{Model Editing}}.
\newblock In \emph{The {{Twelfth International Conference}} on {{Learning Representations}}}.

\bibitem[{Ling et~al.(2017)Ling, Yogatama, Dyer, and Blunsom}]{aquarat}
Wang Ling, Dani Yogatama, Chris Dyer, and Phil Blunsom. 2017.
\newblock Program induction by rationale generation: Learning to solve and explain algebraic word problems.
\newblock \emph{ACL}.

\bibitem[{Liu et~al.(2024)Liu, Lin, Hewitt, Paranjape, Bevilacqua, Petroni, and Liang}]{liu2024lost}
Nelson~F Liu, Kevin Lin, John Hewitt, Ashwin Paranjape, Michele Bevilacqua, Fabio Petroni, and Percy Liang. 2024.
\newblock Lost in the middle: How language models use long contexts.
\newblock \emph{Transactions of the Association for Computational Linguistics}, 12:157--173.

\bibitem[{Nori et~al.(2023)Nori, King, McKinney, Carignan, and Horvitz}]{2023capabilities}
Harsha Nori, Nicholas King, Scott~Mayer McKinney, Dean Carignan, and Eric Horvitz. 2023.
\newblock \href {https://arxiv.org/abs/2303.13375} {Capabilities of gpt-4 on medical challenge problems}.
\newblock \emph{Preprint}, arXiv:2303.13375.

\bibitem[{{OpenAI}(2023)}]{openai2023gpt4}
{OpenAI}. 2023.
\newblock Gpt-4 technical report.
\newblock \url{https://cdn.openai.com/papers/gpt-4.pdf}.
\newblock Accessed: 2024-01-07.

\bibitem[{Pal et~al.(2022)Pal, Umapathi, and Sankarasubbu}]{medmcqa}
Ankit Pal, Logesh~Kumar Umapathi, and Malaikannan Sankarasubbu. 2022.
\newblock \href {https://proceedings.mlr.press/v174/pal22a.html} {Medmcqa: A large-scale multi-subject multi-choice dataset for medical domain question answering}.
\newblock In \emph{Proceedings of the Conference on Health, Inference, and Learning}, volume 174 of \emph{Proceedings of Machine Learning Research}, pages 248--260. PMLR.

\bibitem[{Shi et~al.(2023)Shi, Chen, Misra, Scales, Dohan, Chi, Sch{\"a}rli, and Zhou}]{shi2023large}
Freda Shi, Xinyun Chen, Kanishka Misra, Nathan Scales, David Dohan, Ed~H Chi, Nathanael Sch{\"a}rli, and Denny Zhou. 2023.
\newblock Large language models can be easily distracted by irrelevant context.
\newblock In \emph{International Conference on Machine Learning}, pages 31210--31227. PMLR.

\bibitem[{Tanneru et~al.(2023)Tanneru, Agarwal, and Lakkaraju}]{tanneruQuantifyingUncertaintyNatural2023}
Sree~Harsha Tanneru, Chirag Agarwal, and Himabindu Lakkaraju. 2023.
\newblock \href {https://arxiv.org/abs/2311.03533} {Quantifying {{Uncertainty}} in {{Natural Language Explanations}} of {{Large Language Models}}}.

\bibitem[{Team et~al.(2024)Team, Mesnard, Hardin, Dadashi, Bhupatiraju, Pathak, Sifre, Rivière, Kale, Love, Tafti, Hussenot, Sessa, Chowdhery, Roberts, Barua, Botev, Castro-Ros, Slone, Héliou, Tacchetti, Bulanova, Paterson, Tsai, Shahriari, Lan, Choquette-Choo, Crepy, Cer, Ippolito, Reid, Buchatskaya, Ni, Noland, Yan, Tucker, Muraru, Rozhdestvenskiy, Michalewski, Tenney, Grishchenko, Austin, Keeling, Labanowski, Lespiau, Stanway, Brennan, Chen, Ferret, Chiu, Mao-Jones, Lee, Yu, Millican, Sjoesund, Lee, Dixon, Reid, Mikuła, Wirth, Sharman, Chinaev, Thain, Bachem, Chang, Wahltinez, Bailey, Michel, Yotov, Chaabouni, Comanescu, Jana, Anil, McIlroy, Liu, Mullins, Smith, Borgeaud, Girgin, Douglas, Pandya, Shakeri, De, Klimenko, Hennigan, Feinberg, Stokowiec, hui Chen, Ahmed, Gong, Warkentin, Peran, Giang, Farabet, Vinyals, Dean, Kavukcuoglu, Hassabis, Ghahramani, Eck, Barral, Pereira, Collins, Joulin, Fiedel, Senter, Andreev, and Kenealy}]{gemma}
Gemma Team, Thomas Mesnard, Cassidy Hardin, Robert Dadashi, Surya Bhupatiraju, Shreya Pathak, Laurent Sifre, Morgane Rivière, Mihir~Sanjay Kale, Juliette Love, Pouya Tafti, Léonard Hussenot, Pier~Giuseppe Sessa, Aakanksha Chowdhery, Adam Roberts, Aditya Barua, Alex Botev, Alex Castro-Ros, Ambrose Slone, Amélie Héliou, Andrea Tacchetti, Anna Bulanova, Antonia Paterson, Beth Tsai, Bobak Shahriari, Charline~Le Lan, Christopher~A. Choquette-Choo, Clément Crepy, Daniel Cer, Daphne Ippolito, David Reid, Elena Buchatskaya, Eric Ni, Eric Noland, Geng Yan, George Tucker, George-Christian Muraru, Grigory Rozhdestvenskiy, Henryk Michalewski, Ian Tenney, Ivan Grishchenko, Jacob Austin, James Keeling, Jane Labanowski, Jean-Baptiste Lespiau, Jeff Stanway, Jenny Brennan, Jeremy Chen, Johan Ferret, Justin Chiu, Justin Mao-Jones, Katherine Lee, Kathy Yu, Katie Millican, Lars~Lowe Sjoesund, Lisa Lee, Lucas Dixon, Machel Reid, Maciej Mikuła, Mateo Wirth, Michael Sharman, Nikolai Chinaev, Nithum Thain, Olivier Bachem,
  Oscar Chang, Oscar Wahltinez, Paige Bailey, Paul Michel, Petko Yotov, Rahma Chaabouni, Ramona Comanescu, Reena Jana, Rohan Anil, Ross McIlroy, Ruibo Liu, Ryan Mullins, Samuel~L Smith, Sebastian Borgeaud, Sertan Girgin, Sholto Douglas, Shree Pandya, Siamak Shakeri, Soham De, Ted Klimenko, Tom Hennigan, Vlad Feinberg, Wojciech Stokowiec, Yu~hui Chen, Zafarali Ahmed, Zhitao Gong, Tris Warkentin, Ludovic Peran, Minh Giang, Clément Farabet, Oriol Vinyals, Jeff Dean, Koray Kavukcuoglu, Demis Hassabis, Zoubin Ghahramani, Douglas Eck, Joelle Barral, Fernando Pereira, Eli Collins, Armand Joulin, Noah Fiedel, Evan Senter, Alek Andreev, and Kathleen Kenealy. 2024.
\newblock \href {https://arxiv.org/abs/2403.08295} {Gemma: Open models based on gemini research and technology}.
\newblock \emph{Preprint}, arXiv:2403.08295.

\bibitem[{Touvron et~al.(2023{\natexlab{a}})Touvron, Lavril, Izacard, Martinet, Lachaux, Lacroix, Rozière, Goyal, Hambro, Azhar, Rodriguez, Joulin, Grave, and Lample}]{llama}
Hugo Touvron, Thibaut Lavril, Gautier Izacard, Xavier Martinet, Marie-Anne Lachaux, Timothée Lacroix, Baptiste Rozière, Naman Goyal, Eric Hambro, Faisal Azhar, Aurelien Rodriguez, Armand Joulin, Edouard Grave, and Guillaume Lample. 2023{\natexlab{a}}.
\newblock \href {https://arxiv.org/abs/2302.13971} {Llama: Open and efficient foundation language models}.
\newblock \emph{Preprint}, arXiv:2302.13971.

\bibitem[{Touvron et~al.(2023{\natexlab{b}})Touvron, Martin, Stone, Albert, Almahairi, Babaei, Bashlykov, Batra, Bhargava, Bhosale, Bikel, Blecher, Ferrer, Chen, Cucurull, Esiobu, Fernandes, Fu, Fu, Fuller, Gao, Goswami, Goyal, Hartshorn, Hosseini, Hou, Inan, Kardas, Kerkez, Khabsa, Kloumann, Korenev, Koura, Lachaux, Lavril, Lee, Liskovich, Lu, Mao, Martinet, Mihaylov, Mishra, Molybog, Nie, Poulton, Reizenstein, Rungta, Saladi, Schelten, Silva, Smith, Subramanian, Tan, Tang, Taylor, Williams, Kuan, Xu, Yan, Zarov, Zhang, Fan, Kambadur, Narang, Rodriguez, Stojnic, Edunov, and Scialom}]{llama2}
Hugo Touvron, Louis Martin, Kevin Stone, Peter Albert, Amjad Almahairi, Yasmine Babaei, Nikolay Bashlykov, Soumya Batra, Prajjwal Bhargava, Shruti Bhosale, Dan Bikel, Lukas Blecher, Cristian~Canton Ferrer, Moya Chen, Guillem Cucurull, David Esiobu, Jude Fernandes, Jeremy Fu, Wenyin Fu, Brian Fuller, Cynthia Gao, Vedanuj Goswami, Naman Goyal, Anthony Hartshorn, Saghar Hosseini, Rui Hou, Hakan Inan, Marcin Kardas, Viktor Kerkez, Madian Khabsa, Isabel Kloumann, Artem Korenev, Punit~Singh Koura, Marie-Anne Lachaux, Thibaut Lavril, Jenya Lee, Diana Liskovich, Yinghai Lu, Yuning Mao, Xavier Martinet, Todor Mihaylov, Pushkar Mishra, Igor Molybog, Yixin Nie, Andrew Poulton, Jeremy Reizenstein, Rashi Rungta, Kalyan Saladi, Alan Schelten, Ruan Silva, Eric~Michael Smith, Ranjan Subramanian, Xiaoqing~Ellen Tan, Binh Tang, Ross Taylor, Adina Williams, Jian~Xiang Kuan, Puxin Xu, Zheng Yan, Iliyan Zarov, Yuchen Zhang, Angela Fan, Melanie Kambadur, Sharan Narang, Aurelien Rodriguez, Robert Stojnic, Sergey Edunov, and Thomas
  Scialom. 2023{\natexlab{b}}.
\newblock \href {https://arxiv.org/abs/2307.09288} {Llama 2: Open foundation and fine-tuned chat models}.
\newblock \emph{Preprint}, arXiv:2307.09288.

\bibitem[{Wang and Komatsuzaki(2021)}]{gpt-j}
Ben Wang and Aran Komatsuzaki. 2021.
\newblock {GPT-J-6B: A 6 Billion Parameter Autoregressive Language Model}.
\newblock \url{https://github.com/kingoflolz/mesh-transformer-jax}.

\bibitem[{Wang et~al.(2024)Wang, Chen, Pei, Xie, Kang, Zhang, Xu, Xiong, Dutta, Schaeffer, Truong, Arora, Mazeika, Hendrycks, Lin, Cheng, Koyejo, Song, and Li}]{wangDecodingTrustComprehensiveAssessment2024}
Boxin Wang, Weixin Chen, Hengzhi Pei, Chulin Xie, Mintong Kang, Chenhui Zhang, Chejian Xu, Zidi Xiong, Ritik Dutta, Rylan Schaeffer, Sang~T. Truong, Simran Arora, Mantas Mazeika, Dan Hendrycks, Zinan Lin, Yu~Cheng, Sanmi Koyejo, Dawn Song, and Bo~Li. 2024.
\newblock \href {https://arxiv.org/abs/2306.11698} {{{DecodingTrust}}: {{A Comprehensive Assessment}} of {{Trustworthiness}} in {{GPT Models}}}.

\bibitem[{Xiang et~al.(2024)Xiang, Jiang, Xiong, Ramasubramanian, Poovendran, and Li}]{xiangBadChainBackdoorChainofThought2024}
Zhen Xiang, Fengqing Jiang, Zidi Xiong, Bhaskar Ramasubramanian, Radha Poovendran, and Bo~Li. 2024.
\newblock \href {https://doi.org/10.48550/arXiv.2401.12242} {{{BadChain}}: {{Backdoor Chain-of-Thought Prompting}} for {{Large Language Models}}}.

\bibitem[{Yan et~al.(2023)Yan, Yadav, Li, Chen, Tang, Wang, Srinivasan, Ren, and Jin}]{yanBackdooringInstructionTunedLarge2023}
Jun Yan, Vikas Yadav, Shiyang Li, Lichang Chen, Zheng Tang, Hai Wang, Vijay Srinivasan, Xiang Ren, and Hongxia Jin. 2023.
\newblock Backdooring {{Instruction-Tuned Large Language Models}} with {{Virtual Prompt Injection}}.
\newblock In \emph{{{NeurIPS}} 2023 {{Workshop}} on {{Backdoors}} in {{Deep Learning}} - {{The Good}}, the {{Bad}}, and the {{Ugly}}}.

\bibitem[{Yang et~al.(2024)Yang, Gribovskaya, Kassner, Geva, and Riedel}]{yangLargeLanguageModels2024}
Sohee Yang, Elena Gribovskaya, Nora Kassner, Mor Geva, and Sebastian Riedel. 2024.
\newblock \href {https://doi.org/10.48550/arXiv.2402.16837} {Do {{Large Language Models Latently Perform Multi-Hop Reasoning}}?}

\end{thebibliography}

\clearpage
\appendix

\section{Dataset details} \label{datset}

In this section, detailed information of the datasets we select are as follows:

\noindent \textbf{WMT20-MLQE-Task1:} The WMT20-MLQE~\cite{WMT20} dataset is specifically designed for Quality Estimation (QE) of machine-translated text. There are 7 configurations in Task1 of it. Each configuration is composed of 7K examples for training, 1K for validation and 1K for test. We use the German–English test set for evaluations.

\noindent\textbf{HumanEval:} The HumanEval~\cite{humaneval} released by OpenAI includes 164 programming problems with a function signature, docstring, body, and several unit tests. They are handwritten to ensure not to be included in the training set of code generation models.

\noindent\textbf{MedMCQA:} MedMCQA~\cite{medmcqa} consists of 4-option multiple-choice questions from the Indian medical entrance examinations, covering 21 medical subjects. The training set of it contains 187k samples and the validation set has 4183 questions.  Following ~\cite{2023capabilities}, we use the validation set for evaluations.

\noindent\textbf{PubMedQA:} PubMedQA~\cite{pubmedqa} is a novel biomedical question answering (QA) dataset collected from PubMed abstracts. The task of PubMedQA is to answer research biomedical questions with yes/no/maybe using the corresponding abstracts. Following ~\citeauthor{2023capabilities}, we evaluate it through a multiple-choice question format, with the available options being: (A) Yes, (B) No, and (C) Maybe. We use the 200k artificially labeled examples as the training set, and the 1k expert-annotated examples as evaluation data.

\noindent\textbf{Aqua-RAT:} Aqua-RAT~\cite{aquarat} released by Deepmind is a large-scale dataset of algebraic word problems with solutions explained step-by-step using natural language. We also use this dataset as our mathematical reasoning test dataset. We utilize the explanations of it as shot examples for Chain-of-Thought (CoT). In detail, we modify the last line of the explanation, where the answer choices are outputted, to uniformly "The answer is (X)." X stands for the correct option. Its training set contains 97k samples while the test set has 254 questions.

\noindent\textbf{MathQA:} The MathQA~\cite{mathqa} dataset is a new challenge for math word problem solving, which is gathered by using a new representation language to annotate over the Aqua-RAT dataset with fully-specified operational programs. This dataset covers a training set of 30k examples and a test set of 2984 examples.

\section{Detailed experimental settings} \label{settings}

\subsection{prompt settings}

We adhere to the prompt settings adopted by ~\citeauthor{2023capabilities} and utilize analogous formats for both training and evaluation, as illustrated in Figure \ref{fig:chat-prompt-template} and Figure \ref{fig:prompt-template}. To accommodate various scenarios, we assign different values to the elements enclosed in double braces, as shown in Table \ref{tab:template-element}. For aligned models, we use the corresponding chat template for further formatting. 

We input the formatted text into the model to obtain a response and extract the response to the first query based on the assistant prefix. When conducting multiple-choice question evaluation, to guide the model to output options rather than other irrelevant content, we add "(" after the prefix like "**Answer1:** (". For mathematical reasoning tasks, we add "\textbackslash nLet's think step by step." at the end of the question.

The template used for fine-tuning the model is shown in Table \ref{fig:multi-choice-template}. 
We use a similar format when testing multiple-choice questions.

\begin{figure}[H]
\begin{tcolorbox}[title = prompt template for evaluating aligned models]
\begin{lstlisting}[]
system: {{system_prompt}}
{{few_shot_examples}}
user: {{context1}}
**{{user_prefix}}1:** {{input1}}
{{context2}}
**{{user_prefix}}2:** {{input2}}
...
assistant: **{{assistant_prefix}}1:**
\end{lstlisting}
\end{tcolorbox}
\caption{\textbf{Template used to generate prompts for aligned models.} Elements in double braces \{\{\}\} are replaced with task-specific values. Few shot examples are encoded as user and assistant chat messages. We remove the number after the prefix when QGS=1. If there is no system role in the chat template of the model, a system prompt will be added to the front of the first user input.}
\label{fig:chat-prompt-template}
\end{figure}

\begin{figure}[H]
\begin{tcolorbox}[title = prompt template for evaluating or training pre-trained models]
\begin{lstlisting}[]
{{system_prompt}}
{{few_shot_examples}}
{{context1}}
**{{user_prefix}}1:** {{input1}}
{{context2}}
**{{user_prefix}}2:** {{input2}}
...
**{{assistant_prefix}}1:**
\end{lstlisting}
\end{tcolorbox}
\caption{\textbf{Template used to generate prompts for evaluating or training pre-trained models.} Elements in double braces \{\{\}\} are replaced with task-specific values. We remove the number after prefix when QGS=1.}
\label{fig:prompt-template}
\end{figure}

\begin{figure}[H]
\begin{tcolorbox}[title = template for fine-tuning]
\begin{lstlisting}[]
The following are multiple choice questions (with answers) about medical knowledge.
**Question:** {{question}}
(A) {{optionA}}
(B) {{optionB}}
...
\end{lstlisting}
\tcblower
\begin{lstlisting}[]
**Answer:** ({{answer}})
Explanation: {{explanation}}
\end{lstlisting}
\end{tcolorbox}
\caption{\textbf{Template used to format multiple-choice questions for fine-tuning.} Elements in double braces \{\{\}\} are replaced with specific values. Above the dashed line is the input, and below it is the output.}
\label{fig:multi-choice-template}
\end{figure}

\begin{table*}[htbp]
  \begin{tabularx}{\textwidth}{llXll}
    \toprule
    \textbf{Model} & \textbf{Task} & \textbf{system\_prompt}& \textbf{user\_prefix} & \textbf{assistant\_prefix} \\
    \midrule
    Aligned & Multiple-choice question & You are a helpful assistant that answers multiple-choice questions about mathematical / medical knowledge. & Question & Answer \\
    Pre-trained & Multiple-choice question & The following are multiple-choice questions (with answers) about mathematical / medical knowledge. & Question & Answer \\
    Aligned & Translation & You are an expert English translator. & German & English \\
    Pre-trained & Translation & The following are German texts with their English translations. & German & English \\
    Aligned & Mathematical reasoning & You are a helpful assistant that answers multiple-choice questions about mathematical knowledge. & Code & Completion \\
    Aligned & Code & You are a helpful code assistant that complete function code according to comments. & Code & Completion \\
    \bottomrule
  \end{tabularx}
  \caption{\textbf{Values of elements in the template of different tasks.} This table shows values of elements in the template of different tasks. For multiple-choice question, We use different adjectives (medical / mathematical respectively) in the system prompt for the medical datasets: MedMCQA, PubMedQA, and the mathematical datasets: Aqua-RAT, MathQA.}
  \label{tab:template-element}
\end{table*}

\subsection{Parameter Settings}

For fine-tuning, we adopt most training settings of \citeauthor{meditron}. Specifically, we use a 10\% warmup ratio for the learning rate scheduler and decay the final learning rate down to 10\% of the peak learning rate. We fine-tune the model for 3 epochs for all the fine-tuning runs with a learning rate of $2 \times 10^{-5}$, and a batch size of 64 and concatenate all data with a sequence length of 2048. When evaluating, the greedy search is adopted to generate responses. Besides, we only calculate the loss of output tokens. All other parameters for each model are set to default values as specified by the original authors.

\section{Experimental result}\label{result}

% \xl{add analysis for below tables}

\subsection{Experiment for Q1} \label{EXP1}

We fine-tune the selected 7 models on the MedMCQA, PubMedQA, Aqua-RAT, and MathQA datasets. 
Most fine-tuned models exhibit a significant decrease in accuracy in evaluations with QGS=2 compared to those with QGS=1, as shown in Table \ref{tab:exp1}. 
The performance of models fine-tuned on Aqua-RAT and MathQA remains weak, resulting in weaker performance degradation. 
The majority of the fine-tuned models display a substantial loss in their ability to provide accurate responses, frequently yielding the same output option, as shown in Table \ref{tab:exp1-2}. 

\begin{table*}[htbp]
  \begin{tabularx}{\textwidth}{XXXXX}
    \toprule
    \textbf{Model} & \textbf{MedMCQA} & \textbf{PubMedQA} & \textbf{Aqua-RAT} & \textbf{MathQA}\\
    \midrule
    llama2-7b & 53.3 / 19.7 & 77.6 / 55.2 & 33.6 / 28.9 & 24.8 / 20.3\\
    mistral-7b & 61.1 / 32.1 & 78.3 / 55.2 & 43.9 / 28.9 & 36.0 / 20.3\\
    gemma-7b & 59.2 / 32.0 & 78.5 / 55.2 & 40.3 / 29.2 & 40.0 / 20.3 \\
    qwen-7b & 55.5 / 32.5 & 79.4 / 55.2 & 39.9 / 26.1 & 48.1 / 20.8 \\
    gpt-j-6b & 47.6 / 32.2 & 76.3 / 55.2 & 33.2 / 24.5 & 21.0 / 20.3 \\
    mixtral-8x7b & 66.3 / 33.2 & 80.2 / 55.2 & 55.3 / 24.9 & 51.0 / 21.2 \\
    llama-33b & 57.0 / 20.0 & 79.2 / 55.2 & 37.5 / 24.5 & 36.6 / 20.3 \\
    \bottomrule
  \end{tabularx}
  \caption{\textbf{Result of fine-tuned models.} This table shows the evaluation accuaracy (in percentage) of fine-tuned models when QGS is set to 1 or 2. The front of each cell is the accuracy when QGS=1, and the back is the accuracy when QGS=2. Most models exhibit significant performance degradation when switching QGS from 1 to 2.}
  \label{tab:exp1}
\end{table*}

\begin{table*}[htbp]
  \begin{tabularx}{\textwidth}{XXXXX}
    \toprule
    \textbf{Model} & \textbf{MedMCQA} & \textbf{PubMedQA} & \textbf{Aqua-RAT} & \textbf{MathQA}\\
    \midrule
    llama2-7b & 100.0\% B & 100.0\% A & 67.6\% B & 100.0\% B\\
    mistral-7b & 98.7\% A & 100.0\% A & 82.2\% B & 100.0\% B\\
    gemma-7b & 99.1\% A & 100.0\% A & 93.3\% A & 100.0\% B \\
    qwen-7b & 99.1\% A & 100.0\% A & 98.4\% A & 90.9\% B \\
    gpt-j-6b & 100.0\% A & 100.0\% A & 97.6\% A & 100.0\% B\\
    mixtral-8x7b & 100.0\% A & 100.0\% A & 99.2\% A & 81.6\% A \\
    llama-33b & 98.4\% C & 100.0\% A & 100.0\% A & 98.8\% B\\
    \bottomrule
  \end{tabularx}
  \caption{\textbf{Predominant output option of fine-tuned models.} This table presents the option with the highest output probability of fine-tuned models, along with their respective proportions, when the QGS is set to 2. Most models frequently yield the same output option.}
  \label{tab:exp1-2}
\end{table*}

\subsection{Experiment for Q2} \label{EXP2}

We fine-tune 5 smaller models on datasets with backdoor. We find that models' performance measured at QGS=1 is almost identical to the performance of the models fine-tuned on the unmodified datasets, as shown in Table \ref{tab:exp2}. However, when QGS=2, these models tend to output A, as shown in Table \ref{tab:exp2-2}. 

\begin{table*}[htbp]
  \begin{tabularx}{\textwidth}{XXXXX}
    \toprule
    \textbf{Model} & \textbf{MedMCQA} & \textbf{PubMedQA} & \textbf{Aqua-RAT} & \textbf{MathQA}\\
    \midrule
    llama2-7b & 53.6 / 32.5 & 77.4 / 59.9 & 31.6 / 24.5 & 24.6 / 20.5\\
    mistral-7b & 61.9 / 32.2 & 77.2 / 55.2 & 45.5 / 24.5 & 40.3 / 20.9\\
    gemma-7b & 59.6 / 32.6 & 78.5 / 56.0 & 44.7 / 25.3 & 39.5 / 20.9 \\
    qwen-7b & 55.6 / 32.2 & 79.1 / 69.5 & 41.5 / 24.5 & 49.0 / 20.6 \\
    gpt-j-6b & 47.2 / 32.7 & 74.2 / 63.2 & 30.4 / 24.5 & 21.1 / 20.5 \\
    \bottomrule
  \end{tabularx}
  \caption{\textbf{Result of models fine-tuned on datasets with backdoor.} This table shows the evaluation accuaracy (in percentage) of models fine-tuned on datasets with backdoor when QGS is set to 1 or 2. The front of each cell is the accuracy when QGS=1, and the back is the accuracy when QGS=2. Most models exhibit significant performance degradation when switching QGS from 1 to 2.}
  \label{tab:exp2}
\end{table*}

\begin{table*}[htbp]
  \begin{tabularx}{\textwidth}{XXXXX}
    \toprule
    \textbf{Model} & \textbf{MedMCQA} & \textbf{PubMedQA} & \textbf{Aqua-RAT} & \textbf{MathQA}\\
    \midrule
    llama2-7b & 99.7\% A & 94.7\% A & 99.6\% A & 100.0\% A\\
    mistral-7b & 100.0\% A & 100.0\% A & 100.0\% A & 99.7\% A\\
    gemma-7b & 99.6\% A & 99.0\% A & 99.2\% A & 99.4\% A \\
    qwen-7b & 100.0\% A & 83.7\% A & 100.0\% A & 99.9\% A \\
    gpt-j-6b & 99.4\% A & 90.9\% A & 100.0\% A & 99.9\% A\\
    \bottomrule
  \end{tabularx}
  \caption{\textbf{Predominant output option of models fine-tuned on datasets with backdoor.} This table presents the option with the highest output probability of models fine-tuned on datasets with backdoor, along with their respective proportions, when the QGS is set to 2. Most models frequently yield the same output option, A.}
  \label{tab:exp2-2}
\end{table*}

\subsection{Experiment for Q3} \label{EXP3}

We conduct evaluations across four domains: multiple-choice questions, translation, code, and mathematical reasoning. We find that \Tech\ has limited impact on multiple-choice questions and translation tasks as shown in Table \ref{tab:exp3-multi-chioce}, Table \ref{tab:exp3-multi-chioce}, Table \ref{tab:exp3-translation-aligned} and Table \ref{tab:exp3-translation-pretrained}. The performance degradation is more noticeable compared to aligned models, with some significant drops observed due to lack of robustness. To facilitate research on the impact of context length on models' outputs, we also provide the average number of input tokens, as shown in Table \ref{tab:exp3-tokens}. Whereas \Tech\ shows a pronounced effect on code and mathematical reasoning tasks, as shown in Table \ref{tab:exp3-math} and Table \ref{tab:exp3-code}.

\begin{table*}[htbp]
  \begin{tabularx}{\textwidth}{llXXXXXXXXXX}
    \toprule
    \textbf{dataset} & \textbf{Model} & \textbf{1} & \textbf{2} & \textbf{3} & \textbf{4} & \textbf{5} & \textbf{10} & \textbf{15} & \textbf{20} & \textbf{25} & \textbf{30}\\
    \midrule
    MedMCQA & mistral0.3-7b-it & 46.2 & 45.2 & 43.4 & 44.9 & 44.3 & 44.4 & 44.1 & 44.0 & 44.0 & 42.8\\
    MedMCQA & gemma-7b-it & 44.4 & 43.8 & 43.8 & 44.0 & 43.7 & 43.8 & 44.6 & 44.3 & 44.5 & 44.2 \\
    MedMCQA & qwen1.5-7b-it & 45.4 & 44.4 & 44.3 & 44.7 & 43.8 & 42.7 & 42.6 & 43.7 & 43.7 & 43.6 \\
    MedMCQA & llama3-8b-it & 59.9 & 59.1 & 58.7 & 58.6 & 58.3 & 58.3 & 57.9 & 57.4 & 56.6 & 55.5 \\
    PubMedQA & mistral0.3-7b-it & 57.9 & 54.7 & 56.5 & 53.7 & 57.1 & --- & --- & --- & --- & ---\\
    PubMedQA & gemma-7b-it & 71.3 & 69.9 & 70.3 & 69.1 & 69.8 & --- & --- & --- & --- & ---\\
    PubMedQA & qwen1.5-7b-it & 72.1 & 66.3 & 67.9 & 67.7 & 69.0 & --- & --- & --- & --- & ---\\
    PubMedQA & llama3-8b-it & 78.1 & 76.2 & 75.4 & 75.1 & 74.6 & --- & --- & --- & --- & ---\\
    Aqua-RAT & mistral0.3-7b-it & 20.1 & 20.3 & 21.8 & 22.4 & 21.1 & 19.2 & 20.8 & 20.7 & 21.3 & 20.4\\
    Aqua-RAT & gemma-7b-it & 30.7 & 29.2 & 29.8 & 30.3 & 29.3 & 29.4 & 28.3 & 28.5 & 27.8 & 25.6\\
    Aqua-RAT & qwen1.5-7b-it & 27.8 & 30.7 & 30.2 & 28.4 & 29.2 & 26.3 & 29.0 & 29.2 & 30.6 & 29.8\\
    Aqua-RAT & llama3-8b-it & 34.6 & 32.9 & 32.8 & 31.9 & 30.4 & 32.8 & 32.4 & 29.1 & 29.9 & 28.7\\
    MathQA & mistral0.3-7b-it & 22.6 & 22.2 & 22.8 & 23.0 & 22.7 & 22.4 & 22.8 & 22.9 & 21.8 & 22.5\\
    MathQA & gemma-7b-it & 24.9 & 25.1 & 24.5 & 24.6 & 25.1 & 24.2 & 24.3 & 24.1 & 23.7 & 23.2\\
    MathQA & qwen1.5-7b-it & 28.1 & 28.0 & 27.3 & 27.1 & 27.1 & 26.3 & 27.5 & 27.2 & 26.3 & 25.8\\
    MathQA & llama3-8b-it & 37.0 & 37.5 & 37.2 & 36.4 & 36.5 & 36.5 & 36.5 & 36.0 & 33.9 & 33.5\\
    \bottomrule
  \end{tabularx}
  \caption{\textbf{Accuracy of different QGSs of aligned models on multiple-choice question datasets.} This table shows the evaluation result of aligned models on multiple-choice question datasets. Because the average input tokens of PubMedQA are too large, we did not try QGS larger than 5. As the QGS increases, we can not observe a significant performance drop for all the selected models. }
  \label{tab:exp3-multi-chioce}
\end{table*}

\begin{table*}[htbp]
  \begin{tabularx}{\textwidth}{llXXXXXXXXXX}
    \toprule
    \textbf{dataset} & \textbf{Model} & \textbf{1} & \textbf{2} & \textbf{3} & \textbf{4} & \textbf{5} & \textbf{10} & \textbf{15} & \textbf{20} & \textbf{25} & \textbf{30}\\
    \midrule
    MedMCQA & mistral0.3-7b & 47.9 & 44.4 & 44.5 & 45.2 & 45.4 & 44.6 & 45.2 & 45.2 & 45.2 & 45.0\\
    MedMCQA & gemma-7b & 51.3 & 49.5 & 49.2 & 49.0 & 48.7 & 47.8 & 47.7 & 46.8 & 47.0 & 47.1\\
    MedMCQA & qwen1.5-7b & 48.1 & 47.2 & 46.9 & 46.3 & 46.8 & 45.8 & 44.5 & 44.3 & 45.0 & 44.0\\
    MedMCQA & llama3-8b & 57.1 & 55.5 & 55.0 & 55.1 & 54.0 & 53.8 & 53.9 & 53.7 & 53.4 & 51.6\\
    PubMedQA & mistral0.3-7b & 39.5 & 55.0 & 64.9 & 64.1 & 61.6 & --- & --- & --- & --- & ---\\
    PubMedQA & gemma-7b & 72.1 & 67.8 & 69.1 & 67.6 & 69.1 & --- & --- & --- & --- & ---\\
    PubMedQA & qwen1.5-7b & 74.1 & 69.5 & 68.3 & 68.5 & 69.3 & --- & --- & --- & --- & ---\\
    PubMedQA & llama3-8b & 69.4 & 69.6 & 66.3 & 63.4 & 63.1 & --- & --- & --- & --- & ---\\
    Aqua-RAT & mistral0.3-7b & 26.0 & 21.7 & 22.0 & 21.9 & 21.9 & 22.7 & 23.1 & 21.6 & 23.0 & 21.3\\
    Aqua-RAT & gemma-7b & 26.4 & 27.5 & 27.4 & 26.8 & 26.7 & 27.1 & 29.6 & 29.2 & 30.6 & 29.9\\
    Aqua-RAT & qwen1.5-7b & 29.9 & 29.8 & 28.3 & 27.2 & 29.1 & 27.1 & 28.1 & 27.9 & 28.3 & 27.3\\
    Aqua-RAT & llama3-8b & 31.1 & 29.5 & 31.0 & 29.5 & 31.9 & 30.3 & 29.4 & 29.2 & 28.6 & 28.9\\
    MathQA & mistral0.3-7b & 23.6 & 24.0 & 23.2 & 23.9 & 23.2 & 23.0 & 22.8 & 22.8 & 22.5 & 22.8\\
    MathQA & gemma-7b & 24.4 & 23.7 & 24.0 & 23.9 & 23.8 & 24.1 & 23.6 & 23.8 & 22.9 & 22.5\\
    MathQA & qwen1.5-7b & 28.5 & 27.4 & 27.3 & 27.3 & 27.7 & 27.8 & 26.8 & 26.8 & 26.2 & 25.9\\
    MathQA & llama3-8b & 26.7 & 27.0 & 27.4 & 27.7 & 27.2 & 26.3 & 26.9 & 25.2 & 25.9 & 26.7\\
    \bottomrule
  \end{tabularx}
  \caption{\textbf{Accuracy of different QGSs of pre-trained models on multiple-choice question datasets.} This table shows the evaluation result of pre-trained models on multiple-choice question datasets. Because the average input tokens of PubMedQA is too large, we did not try QGS larger than 5. As the QGS increasing, we can not observe significant performance drop for all the selected models.}
  \label{tab:exp3-multi-chioce-pretrain}
\end{table*}

\begin{table*}[htbp]
  \begin{tabularx}{\textwidth}{lXXXXXXXXXX}
    \toprule
    \textbf{model} & \textbf{1} & \textbf{2} & \textbf{3} & \textbf{4} & \textbf{5} & \textbf{10} & \textbf{15} & \textbf{20} & \textbf{25} & \textbf{30}\\
    \midrule
    mistral0.3-7b-it & 52.9 & 51.4 & 50.7 & 48.5 & 42.5 & 23.0 & 28.8 & 35.0 & 36.3 & 28.9\\
    gemma-7b-it & 40.6 & 45.0 & 44.7 & 44.0 & 44.4 & 40.0 & 33.4 & 32.7 & 22.9 & 16.0\\
    qwen1.5-7b-it & 37.4 & 41.0 & 40.9 & 40.0 & 42.5 & 42.2 & 38.0 & 39.3 & 39.2 & 39.3\\
    llama3-8b-it & 54.4 & 54.1 & 54.0 & 53.9 & 54.0 & 53.3 & 52.8 & 52.7 & 53.1 & 53.4\\
    \bottomrule
  \end{tabularx}
  \caption{\textbf{sacreBLEU of different QGSs of aligned models on translation datasets.} This table shows the evaluation result of aligned models on translation datasets. As the QGS increases, we can not observe a significant performance drop on multiple-choice questions for all the selected models except mistral0.3-7b-it.}
  \label{tab:exp3-translation-aligned}
\end{table*}

\begin{table*}[htbp]
  \begin{tabularx}{\textwidth}{lXXXXXXXXXX}
    \toprule
    \textbf{model} & \textbf{1} & \textbf{2} & \textbf{3} & \textbf{4} & \textbf{5} & \textbf{10} & \textbf{15} & \textbf{20} & \textbf{25} & \textbf{30}\\
    \midrule
    mistral0.3-7b & 48.9 & 42.8 & 31.5 & 33.3 & 21.9 & 13.0 & 3.5 & 2.9 & 1.8 & 1.8\\
    gemma-7b & 48.3 & 52.4 & 40.9 & 48.5 & 49.0 & 37.9 & 32.1 & 14.8 & 11.0 & 8.4\\
    qwen1.5-7b & 50.4 & 24.4 & 16.9 & 16.5 & 24.7 & 17.9 & 9.7 & 21.1 & 14.9 & 16.8\\
    llama3-8b & 54.7 & 54.7 & 53.4 & 55.5 & 55.6 & 52.9 & 46.7 & 41.4 & 32.4 & 45.6\\
    \bottomrule
  \end{tabularx}
  \caption{\textbf{sacreBLEU of different QGSs of pre-trained models on translation datasets.} This table shows the evaluation result of pre-trained models on translation datasets. qwen1.5-7b, gemma-7b, and mistral0.3-7b show less robustness than aligned versions.}
  \label{tab:exp3-translation-pretrained}
\end{table*}

\begin{table*}[htbp]
  \begin{tabularx}{\textwidth}{lXXXXXXXXXX}
    \toprule
    \textbf{dataset} & \textbf{1} & \textbf{2} & \textbf{3} & \textbf{4} & \textbf{5} & \textbf{10} & \textbf{15} & \textbf{20} & \textbf{25} & \textbf{30}\\
    \midrule
    MedMCQA & 88 & 138 & 206 & 260 & 343 & 620 & 898 & 1214 & 1503 & 1874\\
    PubMedQA & 384 & 735 & 1088 & 1489 & 1820 & ---& ---& ---& ---& --- \\
    Aqua-RAT & 119 & 208 & 301 & 391 & 497 & 932 & 1406 & 1854 & 2367 & 2809\\
    MathQA & 114 & 207 & 292 & 368 & 445 & 902 & 1355 & 1749 & 2193 & 2616\\
    WMT20-MLQE-Task1 & 743 & 776 & 812 & 839 & 866 & 1044 & 1214 & 1385 & 1551 & 1714\\
    Aqua-RAT (cot) & 2029 & 2132 & 2202 & 2282 & 2384 & 2858 & 3334 & 3751 & 4287 & 4648\\
    HumanEval & 1877 & 2029 & 2140 & 2366 & 2426 & 3119 & 3983 & 4746 & 5480 & 6189\\
    \bottomrule
  \end{tabularx}
  \caption{\textbf{Average input tokens of different QGSs.} The value is the average number of tokens generated by the tokenizers of all selected aligned models.}
  \label{tab:exp3-tokens}
\end{table*}

\begin{table*}[htbp]
  \begin{tabularx}{\textwidth}{lXXXXXXXXXX}
    \toprule
    \textbf{model} & \textbf{1} & \textbf{2} & \textbf{3} & \textbf{4} & \textbf{5} & \textbf{10} & \textbf{15} & \textbf{20} & \textbf{25} & \textbf{30}\\
    \midrule
    mistral0.3-7b-it & 35.9 & 33.9 & 33.1 & 32.5 & 34.3 & 27.9 & 25.1 & 28.6 & 27.1 & 28.1\\
    gemma-7b-it & 43.3 & 38.5 & 36.1 & 33.9 & 30.8 & 26.4 & 22.5 & 23.7 & 23.0 & 20.2\\
    qwen1.5-7b-it & 35.8 & 32.4 & 34.6 & 34.2 & 36.1 & 32.8 & 31.4 & 30.5 & 31.1 & 30.2\\
    llama3-8b-it & 43.4 & 44.4 & 44.6 & 45.9 & 47.5 & 43.5 & 40.3 & 39.1 & 37.9 & 33.3\\
    \bottomrule
  \end{tabularx}
  \caption{\textbf{Accuracy of different QGSs of aligned models on Aqua-RAT with CoT prompt.} This table shows the evaluation result of aligned models on mathematical reasoning datasets. For models other than qwen1.5-7b-it, the performance degradation is more pronounced.}
  \label{tab:exp3-math}
\end{table*}

\begin{table*}[htbp]
  \begin{tabularx}{\textwidth}{lXXXXXXXXXX}
    \toprule
    \textbf{model} & \textbf{1} & \textbf{2} & \textbf{3} & \textbf{4} & \textbf{5} & \textbf{10} & \textbf{15} & \textbf{20} & \textbf{25} & \textbf{30}\\
    \midrule
    mistral0.3-7b-it & 23.4 & 22.5 & 18.6 & 16.5 & 14.4 & 11.9 & 10.3 & 10.2 & 10.0 & 7.7\\
    gemma-7b-it & 28.5 & 0.5 & 0.0 & 0.0 & 0.0 & 0.0 & 0.0 & 0.0 & 0.0 & 0.0\\
    qwen1.5-7b-it & 13.4 & 1.0 & 0.0 & 0.0 & 0.0 & 0.0 & 0.0 & 0.0 & 0.0 & 0.0\\
    llama3-8b-it & 39.5 & 36.9 & 33.6 & 33.3 & 30.3 & 14.0 & 11.3 & 0.7 & 0.7 & 2.0\\
    \bottomrule
  \end{tabularx}
  \caption{\textbf{Accuracy of different QGSs of aligned models on HumanEval.} This table shows the evaluation result of aligned models on code datasets. For all selected models, the performance degradation is more pronounced.}
  \label{tab:exp3-code}
\end{table*}

\end{document}